**[Article Full Title]**

AI in Proton Therapy Treatment Planning: A Review

**[Short Running Title]**

**[Author Names]**

Yuzhen Ding, PhD[1,*], Hongying Feng, PhD[1,2,3,*], Martin Bues, PhD[1], Mirek Fatyga, PhD[1], Tianming Liu, PhD[4], Thomas J. Whitaker, PhD[5], Haibo Lin, PhD[6], Nancy Y. Lee, MD[7], Charles B. Simone II, MD[6], Samir H. Patel, MD[1], Daniel J. Ma, MD[8], Steven J. Frank, MD[9], Sujay A. Vora, MD[1], Jonathan A. Ashman, MD[1], Wei Liu, PhD[1]

**[Author Institutions]**

[1]Department of Radiation Oncology, Mayo Clinic, Phoenix, AZ 85054, USA

[2]College of Mechanical and Power Engineering, China Three Gorges University, Yichang, Hubei 443002, China

[3]Department of Radiation Oncology, Guangzhou Concord Cancer Center, Guangzhou, Guangdong, 510555, China

[4]School of Computing, the University of Georgia, Athens, GA 30602, USA

[5]Department of Radiation Physics, the University of Texas MD Anderson Cancer Center, Houston, TX 77054, USA

[6]New York Proton Center, New York City, NY 10035, USA

[7]Department of Radiation Oncology, Memorial Sloan Kettering Cancer Center, New York City, NY 10065, USA

[8]Department of Radiation Oncology, Mayo Clinic, Rochester, MN 55905, USA

[9]Division of Radiation Oncology, the University of Texas MD Anderson Cancer Center, Houston, TX 77054, USA

[*]Co-first authors who contribute to this paper equally

**[Corresponding Author Name & Email Address]**

Wei Liu, PhD, e-mail: Liu.Wei@mayo.edu.

**[Author Responsible for Statistical Analysis Name & Email Address]**

Wei Liu, PhD, e-mail: Liu.Wei@mayo.edu.

**[Conflict of Interest Statement for All Authors]**

None

**[Funding Statement]**

This research was supported by NIH/BIBIB R01EB293388, by NIH/NCI R01CA280134, by the Eric & Wendy Schmidt Fund for AI Research & Innovation, and by the Kemper Marley Foundation.

**[Data Availability Statement for this Work]**

Research data are stored in an institutional repository and will be shared upon request to the corresponding author.

**[Acknowledgements]**

None

**Abstract**

**Purpose:** Proton therapy provides superior dose conformity compared to photon therapy, but its treatment planning is challenged by sensitivity to anatomical changes, setup/range uncertainties, and computational complexity. This review evaluates the role of artificial intelligence (AI) in improving proton therapy treatment planning.

**Materials and methods:** Recent studies on AI applications in image reconstruction, image registration, dose calculation, plan optimization, and quality assessment were reviewed and summarized by application domain and validation strategy.

**Results:**

AI has shown promise in automating contouring, enhancing imaging for dose calculation, predicting dose distributions, and accelerating robust optimization. These methods reduce manual workload, improve efficiency, and support more personalized planning and adaptive planning. Limitations include data scarcity, model generalizability, and clinical integration.

**Conclusion:**

AI is emerging as a key enabler of efficient, consistent, and patient-specific proton therapy treatment planning. Addressing challenges in validation and implementation will be essential for its translation into routine clinical practice.

# Introduction

Radiation therapy plays a pivotal role in cancer treatment, offering both curative and palliative benefits by precisely targeting tumors and minimizing unnecessary irradiation to healthy tissues. Among the available modalities, proton therapy has become increasingly attractive for certain cancer disease sites due to its distinct physical characteristics, particularly the Bragg peak and finite range, which allow for highly conformal dose distributions with much smaller exit dose compared to conventional photon therapy. These advantages are especially valuable in pediatric oncology, re-irradiation and cases where tumors are adjacent to critical organs[1-9]. In such cases, minimizing unnecessary dose can potentially reduce long-term toxicities and improve quality of life.

Despite these benefits, proton beams' sensitivities to tissue heterogeneities, anatomical variations, and setup uncertainties[9-12, 13, 14-19] also pose significant challenges for robust treatment delivery. Even small changes in patient anatomy or positioning can lead to substantial deviations between the planned and delivered dose distributions, potentially eroding the clinical benefit of proton therapy[6, 8, 20-31]. Robust optimization strategies have been developed to mitigate these uncertainties, but they often require significant computational time and expert intervention. Furthermore, the integration of multimodal imaging, contouring, and dose calculation into the proton therapy treatment planning process creates additional challenges in efficiency and standardization, which can potentially hinder timely clinical decision-making.

In recent years, artificial intelligence (AI) and deep learning (DL) have emerged as game-changing methods to address these challenges in radiation therapy treatment planning. AI-driven approaches have demonstrated the ability to automate critical components of the radiation therapy treatment planning workflow, including image segmentation/registration, dose prediction, and plan quality evaluation. In proton therapy specifically, DL methods have been developed to accelerate Monte Carlo (MC) dose calculations, enable fast and accurate robust optimization, and provide personalized dose distribution predictions. By reducing manual workload and computational burden, AI has the potential to streamline treatment planning, improve consistency across clinics, and enable more adaptive and patient-specific proton therapy.

This review provides a comprehensive overview of the applications of AI in proton therapy treatment planning. We examine recent advances in 3-dimensional (3D) CT reconstruction, deformable image registration (DIR), auto-segmentation, plan evaluation, dose calculation, plan optimization and adaptive radiation therapy (ART), highlighting how AI has been integrated into each step of the clinical workflow. We also discuss current methodological limitations, regulatory considerations, and emerging research directions, hoping to accelerate the clinical adoption of AI-assisted proton therapy treatment planning.

## A. CT Reconstruction

Computed Tomography (CT) imaging plays a central role in proton therapy treatment planning and delivery. As the primary imaging modality in proton therapy, CT provides detailed anatomical information essential for accurate delineation of tumors and surrounding organs at risk (OARs). Despite anatomical guidance, CT offers Hounsfield Unit (HU) values, which can be converted into relative stopping power (RSP) maps. These RSP maps are critical for calculating the proton beam range and dose distributions.

During the treatment course, repeated on-board imaging (OBI) is commonly used to verify patient positioning and monitor anatomical changes over time. This enables the detection of setup deviations or inter-fractional anatomical variations, which may necessitate the ART process (will be discussed in the ART section). Though dispensable in proton therapy treatment planning, CT has coherent drawbacks that limit its frequent use in clinical practice in a certain treatment course: rather long acquisition time and non-negligible radiation dose once repeatedly used. In current radiotherapy practice, several on-board imaging (OBI) modalities are routinely used, including CT-on-rails (CToR), cone-beam CT (CBCT), and orthogonal kilovoltage (kV) x-ray imaging. CToR provides CT of diagnostic quality but requires transferring the patient from the CT scanner to the treatment isocenter, potentially introducing additional positioning time and error. In contrast, CBCT allows imaging at the treatment isocenter, avoiding patient transfer. Yet, it is more prone to image artifacts and scatter that can degrade image quality. Orthogonal kV imaging offers real-time 2-dimensional (2D) projection images at significantly lower radiation doses and is often used for daily patient setup. However, it only visualizes 2D bony anatomy, restricting its use in treatment planning.

To overcome the limitations of current OBI approaches, AI-based solutions are rapidly emerging and have demonstrated strong potential in enhancing image quality (e.g., denoising, artifact correction). These DL models are capable of generating synthetic CT (sCT) images from lower-dose or lower-quality inputs—such as CBCT, limited-angle CT, or even orthogonal kV x-ray projections—as well as from non-ionizing imaging modalities like magnetic resonance imaging (MRI) and ultrasound.

## A.1 Reconstruction CT from 3D Inputs

In radiotherapy, early studies of applying AI on 3D CT reconstruction often focused on converting CBCT to CT, as CBCT is routinely acquired for image guidance and shares similar geometric characteristics with diagnostic CT. This made it a practical starting point for validating the feasibility of DL-based 3D CT reconstruction. In terms of the model utilized, such approaches can be categorized into two classes in general, i.e., UNet-based and generative adversarial network (GAN)-based. UNet consists of an encoder and a decoder along with additional skip connections to extract and reconstruct multi-scale image representations from input domains to output domains, thus learning to go from CBCT to CT. For example, Chen et al.[32] trained a UNet model to synthesize CT from CBCT, achieving a relatively good mean absolute error (MAE). Other UNet-based CT reconstruction from CBCT can be found in Liu et al[33] and Thumerer et al[34]. For ART, GAN-based models, especially CycleGAN was frequently used[35-38].

There are also frameworks that utilized other DL-based models or combinations of UNet and GAN, for example, the SwinUNETR[39] transformer architecture was applied to CBCT-to-CT translation in prostate cancer, outperforming conventional UNet models in both image quality and dosimetric accuracy—an indication of the growing relevance of transformer-based DL in ART. Other related work can be found as well[40-42]. Furthermore, several studies have explored the reconstruction of 3D CT images from MRI[43-47]. Nevertheless, these approaches are not yet widely adopted in clinical proton therapy due to hardware constraints and uncertainties in dose calculation and are therefore not further discussed in this review.

## A.2 Reconstruction CT from 2D Inputs

Compared with reconstructing 3D CT from volumetric inputs such as CBCT, synthesizing CT from 2D images, although more clinically realistic, is considerably more challenging, as the information preserved in 2D projections is inherently sparse relative to full 3D CT data. To validate feasibility, early studies investigated AI-driven reconstruction of 3D CT from digital reconstructed radiographs (DRRs)[48-54], which are clean, noise-free 2D images generated from planning CT or CToR scans. While conceptually inspiring, such approaches are not clinically applicable, as DRRs themselves require prior 3D imaging for generation. From a practical standpoint, only independently acquired 2D images, such as kV projections, are qualified as meaningful input for 2D-to-3D reconstruction models. Although some studies[55-57] explored X-ray–based training and testing, these images were often synthesized from 3D volumes using ray-tracing, thereby enforcing a strict one-to-one correspondence that did not reflect clinical reality. Recently, however, a framework termed kV2CTConverter[58] was introduced, which directly synthesizes 3D CT from clinically acquired kV images without reliance on prior CT. Results demonstrated both feasibility and near real-time performance, underscoring its use for more accurate 3D patient positioning and potential for ART image guidance.

Despite these advances, several challenges remain before 2D-to-3D CT synthesis can be fully integrated into clinical proton therapy. A key limitation arises from the severe data sparsity of 2D projections, making the reconstruction problem inherently ill-posed and highly sensitive to noise, motion, and anatomical variability. Furthermore, the absence of strict correspondence between kV images and ground-truth CTs complicates supervised training, necessitating reliance on synthetic datasets or registration-based methods that may limit generalizability. Robustness against motion artifacts and anatomical changes is another open concern, particularly given the stringent dosimetric requirements of proton therapy. Future research directions include integrating physics-informed priors such as ray-tracing or MC–based projection operators to constrain the learning process, as well as leveraging multi-view acquisitions and data-efficient learning strategies. Finally, large-scale, multi-institutional validation will be critical to establishing the clinical reliability of these models and to enabling their deployment in real-time adaptive proton therapy workflows.

**B. DIR**

During a certain treatment course, aligning all the initial planning and verifying images to the same framework is the foundation for subsequent image-based procedures. DIR is the process of spatially aligning two or more medical images by estimating a dense, voxel-wise deformation vector field (DVF) that maps anatomical structures from one image to another while accounting for complex, non-rigid changes such as organ motion, deformation, and anatomical variation. Therefore, DIR plays a critical role in ART workflows. We briefly recall the status of topic here and encourage the readers to refer the review[59] for more details.

DL-based DIR approaches can generally be categorized into three classes according to the learning paradigm: unsupervised learning, supervised learning, and learning with joint tasks. Among these, unsupervised learning is the predominant paradigm for DL-based DIR. The unsupervised approach avoids the need for ground-truth DVFs[60-62] by optimizing a loss function based on similarity metrics between the fixed image and the warped image generated from the moving image and the predicted DVF, thereby improving the predictive capability of the model. In contrast, fully supervised approaches require the availability of ground-truth DVFs[63-66]. Joint learning frameworks leverage knowledge from additional tasks or modalities to achieve more comprehensive and accurate DVF predictions.

**B.1 Unsupervised DIR Approaches**

VoxelMorph[60] was among the first DL–based DIR methods and is widely regarded as a foundational baseline for subsequent approaches. It is an unsupervised framework that employs a UNet backbone combined with a Spatial Transformer Network (STN) to apply the predicted DVF to the moving image during inference, enabling rapid DVF prediction. Subsequent studies have enhanced VoxelMorph's performance across various image modalities and anatomical sites by incorporating strategies such as random masking during training[67], probabilistic modeling[68] of DVFs, cycle-consistency losses[69] to stabilize training, and efficient hyperparameter tuning[70].

For large or complex cases, sequential or path-wise registration strategies have been proposed to improve accuracy, which is particularly critical for reliable dose accumulation in ART. Examples include multi-scale registration in LapIRN[71], longitudinal registration in Seq2Morph[72] and others[73-76]. Additionally, some researchers have explored vision transformer (ViT)–based models for DIR, achieving more global and generalizable DVFs[61].

## B.2 Supervised DIR Approaches

Quicksilver[77] is a patch-wise DL framework that predicts Large Deformation Diffeomorphic Metric Mapping (LDDMM) momentum in a supervised manner, retaining diffeomorphic properties while achieving fast inference. Variants include a probabilistic model for uncertainty estimation and a correction network to improve accuracy. Other approaches leverage weak or contour supervision to guide registration: for example, Hu et al.[78] used high-level anatomical labels during training but operated label-free at inference, enabling real-time multimodal (MRI–US) registration with high accuracy.

## B.3 Learning with Joint tasks

Despite these advances, both unsupervised and supervised DL-DIR approaches have limitations. Unsupervised methods, while flexible, can be sensitive to intensity variations, noise, or anatomical inconsistencies, potentially resulting in suboptimal or non-physical deformations. Supervised methods, on the other hand, rely on accurate voxel-level correspondences or high-quality ground-truth DVFs, which are often difficult or impossible to obtain for complex anatomies or multimodal images. These challenges have motivated the development of joint-task or multi-task DL-based DIR frameworks, which simultaneously optimize registration along with related tasks—such as segmentation, contour guidance, or dose accumulation—enhancing accuracy, robustness, and clinical applicability. For example, Liang et al.[79] proposed CT-to-CBCT DIR for head and neck (H&N) auto segmentation, Xie et al.[80] developed an unsupervised GAN-based STN for longitudinal CBCT abdominal RT, Smolders et al.[81] introduced DVF prediction with integrated plausibility scoring for quality assurance (QA), and Hemon et al.[82] demonstrated contour-supervised DIR for prostate CBCT-guided RT.

# C. Contours and Quality Assurance (QA)

## C.1 Contours

The segmentation of targets and OARs is a crucial component of radiation treatment planning. Currently the manual segmentation of these structures is still the standard of care, which is not only tedious and time-consuming but also inevitably prone to inter- and intra-observer variability. Due to the repetitive nature, segmentation tasks constitute ideal candidates for AI-based auto-segmentation. Readers are encouraged to refer to reviews[83, 84] for a more comprehensive understanding on this topic.

Over the past years, AI-based auto-segmentation on common medical images, such as CT, MR, and PET, has rapidly evolved since the introduction of convolutional neural network (CNN), leading to extremely diversified research-orientated models and various commercial solutions. Compared to successful application of AI-based auto-segmentation for OARs, developing strides of AI tools for target delineation fall behind due to the great difficulty in exact voxel-wise differentiation of affected tissue from the surrounding normal tissue. By far, UNet is the most dominating architecture across all disease sites. The hierarchical down-sampling encoder and up-sampling decoder, together with skip connections at each scale forms the U-shape network and enables it to effectively capture low-level and high-level features, leading to accurate segmentation results. The introduction of ViT[85] further enriches the pure CNN encoder network with CNN-transformer hybrid network.[86-88] To further enhance the auto-segmentation performance, hybrid network that combines the traditional model-driven methods guided by explicit anatomical and context information and the data-driven AI-based methods.[89, 90] Dynamic and interactive editing strategy has also been integrated into AI models via reinforcement learning (RL) for iterative contour refinement.[91, 92]

## C.2 QA

With the fast evolution in AI algorithms, auto-segmentation tools have demonstrated remarkable potential for fast and consistent contour delineation. Nonetheless, the accuracy of the delineation still needs to be evaluated (i.e. QA) and approved slice by slice before those contours are adopted for subsequent steps in radiotherapy to achieve optimal and safe treatment. Metrics to evaluate the performance of auto-segmentations generally fall into four categories: intensity and inter-image geometric (such as the most used Dice similarity coefficient and Hausdorff distance) metrics, recorded time savings, subjective scoring, and dosimetric metrics[93, 94], with geometric and intensity metrics being the most employed.

The auto-segmentations can be QAed solely[95-101], or accompanied by another independent set of auto-segmentations[102-105], with the latter one capable of calculating inter-image geometric metrics in the QA process. Thanks to the existence of accompanying auto-segmentations, the evaluation in the latter scenario is more straightforward and mainly utilizes metrics calculation and judgement[102-104] or conventional machine learning (ML)[105]. In contrast, with less information, the evaluation of sole auto-segmentation set is more complicated. A few studies used intra-image geometric metrics of a certain contour or between two contours and intensity metrics for feature selection and ML-based model training[98, 100], while many others exploited the benefit of AI models.[95-97, 99, 101] Rhee *et al.* trained a CNN-based auto-segmentation tool and evaluated its ability for contour error detection for OARs in head and neck patients.[101] Men *et al.*

achieved promising auto-segmentation QA results for lung patients using a deep active learning technique.[95] Chen *et al.* developed a CNN model incorporating probability and uncertainty maps with CT images for auto-segmentation QA for breast cancer.[96] Zhao *et al.* introduced a one-class support vector machine (SVM) to determine the quality of a contour after a ResNet-152 feature extractor.[97] Zarenia *et al.* developed a SegResNet-based QA tool for auto-segmented abdominal contours.[99]

### D. Dose prediction/denoising

Accurate dose calculation is the cornerstone of radiotherapy planning and evaluation. Traditionally, dose distributions are calculated using physics-based algorithms such as pencil beam convolution (PBC)[106-109], collapsed cone convolution (CCC), and MC simulations[110-113]. Among these, MC remains the gold standard, especially in proton therapy, where its ability to model particle interactions and heterogeneities provides unmatched accuracy.

However, the clinical implementation of MC faces major obstacles: computational cost, long runtimes, and storage demands. These barriers are particularly acute in ART, where frequent verification dose calculation and re-planning are required to account for anatomical and physiological changes. Consequently, there is an urgent need for methods that are both fast and accurate, enabling real-time ART workflows while maintaining confidence in dose distributions.

Recent years have seen rapid development of DL methods aimed at overcoming these limitations. Two primary strategies have emerged: dose prediction and dose denoising, where prediction accelerates workflows by approximating physics, while denoising preserves physics accuracy at reduced cost.

## D.1 Dose Prediction Approaches

Dose prediction approaches learn a direct mapping from input images (CT, CBCT, or sCT) and/or structure contours, and/or basic beam information to dose distributions, bypassing iterative physics simulation. Typical architectures include UNet variants[114-117] for voxel-wise regression, GANs[118, 119] for sharper, more realistic dose maps, and more recently, transformers and hybrid models[120-123] such as using ViT, combining ResNet, for long-range dependency modeling. It enjoys extremely fast inference (seconds), thus capable of facilitating efficient decision-makings in initial treatment planning and ART workflows. However, current DL-based models are typically sub-disease/disease-site specific. Therefore, numerous models are needed for different disease sites and/or sub disease sites, which are time-consuming and impractical for routine clinical practices. In addition, the training data are relatively consistent with clear and regular structure contours, which could lead to less robust models and potential model overfit. It is more practical to have a single model trained with data from one disease site without any strictly required input data and that the trained model can be applied to various disease sites without re-training or fine-tuning for routine use. Although there is one work[124] proposed to achieve site-agnostic performance by training on the source site (e.g., prostate) and fine-tuning with minimal effort to a different site (e.g., H&N), the evaluation was very limited.

Furthermore, in the absence of physics-based constraints, such models may suffer from limited physical interpretability. Zhang et al. tried to develop a physics-aware DL-based dose prediction method in proton therapy by adopting a so-called noisy probing dose[125], however, such methods are still scarce. Consequently, an alternative line of research has emerged that investigates dose denoising by utilizing noisy dose distributions as inputs, while simultaneously exploiting the inherent physical information within.

## D. 2 Dose Denoising Approaches

Due to the nature of statistics in MC simulations, decreased noise requires quadratically increased simulating particles, thus quadratically prolonged runtime. Therefore, incorporating sufficient simulating particles to suppress stochastic noise and pursue high-fidelity in MC simulations turn to be less cost-effective. And DL–based denoising has emerged as a powerful alternative, offering substantial noise reduction with minimal loss of accuracy. UNets[126, 127] have been widely adopted to learn mappings between noisy and reference doses, demonstrating improved accuracy compared to conventional filtering. More advanced frameworks, such as GANs[128], have been applied to further enhance structural preservation and mitigate over smoothing. Most recent studies are exploring transformer and diffusion models[129] for improved generalization across patient cohorts and sub-disease/disease sites. Collectively, these approaches enable high-quality and near real-time dose estimation, accelerating the translation of MC-based treatment planning into clinical practice in radiotherapy. The future direction in this regard may include developing hybrid models that combine prediction and denoising (e.g., predicting a first-pass dose, then refining with denoising modules). Moreover, integrating MC kernels or dose deposition models into DL networks for improved interpretability is also essential. In addition, uncertainty quantification and clinical validation are points worth exploring, especially in proton therapy, where prediction errors must be bound to ensure patient safety and multi-institutional trials are needed to prove generalizability and robustness.

## E. Optimization

Influence matrix (IM) is the central concept for conventional inverse treatment planning. By definition, IM quantifies the contribution of each proton beamlet to the dose at each voxel within the selected constrained patient volume of interest (VOI). By tuning the weight of each beamlet, the dose distribution in VOI can be optimized, i.e., uniform and conforming dose in targets and "as low as reasonably achievable" dose in OARs. Robust optimization strategies[130], including worst-case[14, 17, 131, 132] and probabilistic formulations[133-135], have been developed to safeguard against uncertainties in proton therapy by encompass multiple uncertainty scenarios in optimization. However, since one IM corresponds to one uncertainty scenario, IM-based robust optimizations are computationally demanding due to the need to pre-calculate and store large influence matrices. And the conventional "trial and error" iterative optimization process can take long (even hours) to converge. Under such traditional IM-based optimization scheme, the dose calculation process burden can be largely alleviated by the AI-based methods discussed in Section "Dose prediction/denoising".

In addition to traditional optimization workflows, spot weight prediction has emerged as a promising strategy to accelerate treatment planning. By leveraging patient anatomy and desired dose distributions, DL models can directly predict optimal spot weights or fluence maps, by passing some of the computationally intensive steps involved in conventional inverse treatment planning. These predictive approaches can reduce the optimization time for plan of clinically acceptable quality to seconds, providing a natural bridge between accurate dose calculation, influence matrix utilization, and efficient plan generation. This integration is particularly valuable in ART, where rapid plan adaptation is required to accommodate anatomical changes, and near real-time dose re-optimization becomes critical for maintaining treatment efficacy and safety.

Several studies have contributed to treatment plan optimization through DL techniques[136-140], primarily in the context of photon therapy, particularly volumetric modulated arc therapy (VMAT) and intensity modulated radiation therapy (IMRT). These works have demonstrated the feasibility of directly predicting fluence maps or spot weights from patient anatomy and dose objectives. However, due to the fundamental differences in the dose distributions and the physics mechanism beneath, between photons and protons, such DL models developed for photon therapy, though well explored and considerably promising, are not directly transferable to proton therapy and require substantial modifications and rigorous validation. Moreover, several technical barriers remain before DL–based optimization can be fully integrated into routine clinical practice. These include the limited interpretability of purely data-driven models, the incorporation of uncertainty scenario modeling to account for range and motion variations, and the necessity of large-scale, multiinstitutional validation to establish generalizability across diverse patient cohorts and treatment protocols. Looking ahead, the integration of DL with physics-informed constraints and MC–based dose calculation engines represents a compelling direction for advancing fast, accurate, and interpretable optimization frameworks capable of supporting real-time adaptive proton therapy.

## F. ART

During the treatment course of proton therapy, inter-fractional anatomical changes may occur and cause under-treatment of tumors or over-exposure of surrounding OARs that can lead to local recurrence and unexpected treatment-related adverse events. To address the anatomical changes, ART has been introduced[141, 142]. In ART, patients first undergo periodic verification imaging during the treatment course to obtain information about their internal anatomical changes. The clinicians will then assess whether the dose from the initial plan is still within the allowable tolerance based on the periodic verification images, and a re-plan is needed if the initial plan does not meet clinical requirements. The implementation of ART essentially involves all key components (image acquisition, structure segmentation, dose verification, plan optimization, and patient-specific quality assurance (PSQA)) in routine radiotherapy, which brings significant additional clinical workload in aspects of human, equipment, financial, and time. With execution efficiency of ART workflow evolves from the delayed offline mode between fractions, to the timelier online mode before a fraction, to the most sensitive real-time mode during a fraction, the additional workload grows remarkably. Therefore, resource demanding nature of ART workflow sets up an ideal, though challenging, platform for AI tools to make their supportive roles in proton therapy to the fullest.

On one hand, all AI tools developed for routine proton therapy can technically be used in ART proton therapy since the key components of these two workflows are essentially the same, for instance the rather extensively exploited dose prediction[143-145] and structure segmentation[83, 84, 146, 147]. On the other hand, with additional information acquired during the previously delivered fractions, more specific and complicated AI tools can be developed in the ART workflow, as discussed below.

(1) Image acquisition. Currently CT is the more commonly used imaging modality for ART, compared to CBCT (degraded image quality) or MR (lacking information for dose calculation) images. However, once frequently used, CT will introduce non-negligible radiation dose. And the rather long acquisition time is still suboptimal for online ART. AI-based sCT generation has been thoroughly exploited as reviewed in Section "CT Reconstruction".

(2) Image registration. The mapping relationship, i.e. DVF, between the two registered images is the requisite for accumulated dose calculation and contour propagation in ART. The conventional similarity metrics-based iterative optimization process of DIR often takes minutes to converge, which is unaffordable in oART workflows. AI-based solutions have shown appealing potential in this topic as reviewed in Section "DIR".

(3) Spot weight finetuning. Re-optimization is the central component for ART and can be implemented in a way of full re-optimization with a new spot map, which anticipates superior plan quality but is more complicated and time-consuming, or in a way of spot weight finetuning, which is easier at the cost of potentially less improved adaptive plan quality. For the latter way, a literature[148] reported a model that could retuning spot weights based on the original influence matrix using the after-delivery verification dose.

(4) Anatomical change prediction. For the slow inter-fractional anatomical changes, rather than plan adaption after the verification imaging, AI can be used to predict such anatomical changes[149-152] that "adaptive" considerations could be included ahead in the initial plan optimization stage. For the fast intra-fractional anatomical changes, usually in thoracic and abdominal regions, the tumor motion can also be predicted using AI tools for real-time adaption.[153-155]

## Discussion

AI applications in proton therapy span multiple domains, including imaging, dose calculation, plan optimization, adaptive therapy, and outcome prediction. Each area shows encouraging progress, yet several overarching challenges remain. First, many DL models lack physical interpretability, which limits clinical trust and hinders safe deployment. Incorporating physics-informed constraints, such as dose kernels or MC approximations, can improve both robustness and transparency. Second, uncertainty management is particularly critical in proton therapy, where range uncertainties and anatomical changes significantly affect dose distributions; integrating robust optimization strategies with AI-driven predictions remains an open question. Third, the field suffers from limited availability of large, annotated datasets, restricting model generalizability across institutions and patient populations.

Emerging advances in large language models (LLMs) and multi-modal AI offer potential avenues to support proton therapy research and clinical practice[156-161]. While they do not directly address fundamental limitations such as physical interpretability or uncertainty, LLMs may assist in integrating textual clinical records[162], imaging and dosimetric data[163-166], providing more comprehensive decision support[156]. Multi-modal AI approaches that combine CT, MRI and treatment parameters can exploit joint information across different modalities[167], potentially improving predictive accuracy and enabling more personalized treatment planning. These technologies are still exploratory, but their incorporation into hybrid physics–AI frameworks may enhance model utility and guide future clinical translation.

Finally, while early studies demonstrate technical feasibility, few have undergone prospective validation, and regulatory pathways for AI adoption in radiation oncology are still evolving. Future directions should therefore prioritize hybrid physics–AI frameworks, multi-modal data integration, standardized evaluation benchmarks, and multi-institutional collaborations to ensure reproducibility, clinical reliability, and safe deployment of AI-driven solutions.

## Conclusion

AI holds strong potential to transform proton therapy by enabling faster, more accurate, and more adaptive treatment planning and delivery. However, widespread clinical adoption will depend on developing interpretable, physics-aware models and validating them in large, diverse patient cohorts. With continued interdisciplinary collaboration, AI-driven proton therapy can advance toward safer, more personalized cancer care.